 \documentclass[12pt]{amsart}
\usepackage[T1]{fontenc}
\usepackage{amssymb}
\usepackage{a4wide}
\usepackage{graphicx}
\usepackage{verbatim}
\usepackage{amsmath}
\usepackage{version}

 \theoremstyle{plain}    
 \newtheorem{thm}{Theorem}[section]
 \numberwithin{equation}{section} 
 \numberwithin{figure}{section} 
 \theoremstyle{plain}
 \newtheorem{prop}[thm]{Proposition} 
 \theoremstyle{plain}    
 \theoremstyle{plain}    
 \newtheorem{cor}[thm]{Corollary} 
 \theoremstyle{plain}    
\theoremstyle{definition}
\newtheorem{rem}[thm]{Remark} 
\theoremstyle{definition}

\newcommand{\Lim}{\mathop{\longrightarrow}\limits}

\begin{document}
\newcommand{\nwc}{\newcommand}
\nwc{\nwt}{\newtheorem}
\nwt{coro}{Corollary}
\nwt{ex}{Example}


\nwc{\mf}{\mathbf} 
\nwc{\blds}{\boldsymbol} 
\nwc{\ml}{\mathcal} 


\nwc{\lam}{\lambda}
\nwc{\del}{\delta}
\nwc{\Del}{\Delta}
\nwc{\Lam}{\Lambda}
\nwc{\elll}{\ell}

\nwc{\IA}{\mathbb{A}} 
\nwc{\IB}{\mathbb{B}} 
\nwc{\IC}{\mathbb{C}} 
\nwc{\ID}{\mathbb{D}} 
\nwc{\IE}{\mathbb{E}} 
\nwc{\IF}{\mathbb{F}} 
\nwc{\IG}{\mathbb{G}} 
\nwc{\IH}{\mathbb{H}} 
\nwc{\IN}{\mathbb{N}} 
\nwc{\IP}{\mathbb{P}} 
\nwc{\IQ}{\mathbb{Q}} 
\nwc{\IR}{\mathbb{R}} 
\nwc{\IS}{\mathbb{S}} 
\nwc{\IT}{\mathbb{T}} 
\nwc{\IZ}{\mathbb{Z}} 
\def\bbbone{{\mathchoice {1\mskip-4mu {\rm{l}}} {1\mskip-4mu {\rm{l}}}
{ 1\mskip-4.5mu {\rm{l}}} { 1\mskip-5mu {\rm{l}}}}}
\def\bbleft{{\mathchoice {[\mskip-3mu {[}} {[\mskip-3mu {[}}{[\mskip-4mu {[}}{[\mskip-5mu {[}}}}
\def\bbright{{\mathchoice {]\mskip-3mu {]}} {]\mskip-3mu {]}}{]\mskip-4mu {]}}{]\mskip-5mu {]}}}}
\nwc{\setK}{\bbleft 1,K \bbright}
\nwc{\setN}{\bbleft 1,\cN \bbright}


\nwc{\va}{{\bf a}}
\nwc{\vb}{{\bf b}}
\nwc{\vc}{{\bf c}}
\nwc{\vd}{{\bf d}}
\nwc{\ve}{{\bf e}}
\nwc{\vf}{{\bf f}}
\nwc{\vg}{{\bf g}}
\nwc{\vh}{{\bf h}}
\nwc{\vi}{{\bf i}}
\nwc{\vI}{{\bf I}}
\nwc{\vj}{{\bf j}}
\nwc{\vk}{{\bf k}}
\nwc{\vl}{{\bf l}}
\nwc{\vm}{{\bf m}}
\nwc{\vM}{{\bf M}}
\nwc{\vn}{{\bf n}}
\nwc{\vo}{{\it o}}
\nwc{\vp}{{\bf p}}
\nwc{\vq}{{\bf q}}
\nwc{\vr}{{\bf r}}
\nwc{\vs}{{\bf s}}
\nwc{\vt}{{\bf t}}
\nwc{\vu}{{\bf u}}
\nwc{\vv}{{\bf v}}
\nwc{\vw}{{\bf w}}
\nwc{\vx}{{\bf x}}
\nwc{\vy}{{\bf y}}
\nwc{\vz}{{\bf z}}
\nwc{\bal}{\blds{\alpha}}
\nwc{\bep}{\blds{\epsilon}}
\nwc{\barbep}{\overline{\blds{\epsilon}}}
\nwc{\bnu}{\blds{\nu}}
\nwc{\bmu}{\blds{\mu}}



\nwc{\bk}{\blds{k}}
\nwc{\bm}{\blds{m}}
\nwc{\bM}{\blds{M}}
\nwc{\bp}{\blds{p}}
\nwc{\bq}{\blds{q}}
\nwc{\bn}{\blds{n}}
\nwc{\bv}{\blds{v}}
\nwc{\bw}{\blds{w}}
\nwc{\bx}{\blds{x}}
\nwc{\bxi}{\blds{\xi}}
\nwc{\by}{\blds{y}}
\nwc{\bz}{\blds{z}}


\nwc{\cA}{\ml{A}}
\nwc{\cB}{\ml{B}}
\nwc{\cC}{\ml{C}}
\nwc{\cD}{\ml{D}}
\nwc{\cE}{\ml{E}}
\nwc{\cF}{\ml{F}}
\nwc{\cG}{\ml{G}}
\nwc{\cH}{\ml{H}}
\nwc{\cI}{\ml{I}}
\nwc{\cJ}{\ml{J}}
\nwc{\cK}{\ml{K}}
\nwc{\cL}{\ml{L}}
\nwc{\cM}{\ml{M}}
\nwc{\cN}{\ml{N}}
\nwc{\cO}{\ml{O}}
\nwc{\cP}{\ml{P}}
\nwc{\cQ}{\ml{Q}}
\nwc{\cR}{\ml{R}}
\nwc{\cS}{\ml{S}}
\nwc{\cT}{\ml{T}}
\nwc{\cU}{\ml{U}}
\nwc{\cV}{\ml{V}}
\nwc{\cW}{\ml{W}}
\nwc{\cX}{\ml{X}}
\nwc{\cY}{\ml{Y}}
\nwc{\cZ}{\ml{Z}}


\nwc{\tA}{\widetilde{A}}
\nwc{\tB}{\widetilde{B}}
\nwc{\tE}{E^{\vareps}}
\nwc{\tk}{\tilde k}
\nwc{\tN}{\tilde N}
\nwc{\tP}{\widetilde{P}}
\nwc{\tQ}{\widetilde{Q}}
\nwc{\tR}{\widetilde{R}}
\nwc{\tV}{\widetilde{V}}
\nwc{\tW}{\widetilde{W}}
\nwc{\ty}{\tilde y}
\nwc{\teta}{\tilde \eta}
\nwc{\tdelta}{\tilde \delta}
\nwc{\tlambda}{\tilde \lambda}
\nwc{\ttheta}{\tilde \theta}
\nwc{\tvartheta}{\tilde \vartheta}
\nwc{\tPhi}{\widetilde \Phi}
\nwc{\tpsi}{\tilde \psi}

\nwc{\To}{\longrightarrow} 

\nwc{\ad}{\rm ad}
\nwc{\eps}{\epsilon}
\nwc{\ep}{\epsilon}
\nwc{\vareps}{\varepsilon}

\def\ep{\epsilon}
\def\tr{{\rm tr}}
\def\Tr{{\rm Tr}}
\def\i{{\rm i}}
\def\mi{{\rm i}}
\def\e{{\rm e}}
\def\sq2{\sqrt{2}}
\def\sqn{\sqrt{N}}
\def\vol{\mathrm{vol}}
\def\defi{\stackrel{\rm def}{=}}
\def\t2{{\mathbb T}^2}
\def\s2{{\mathbb S}^2}
\def\hn{\mathcal{H}_{N}}
\def\shbar{\sqrt{\hbar}}
\def\A{\mathcal{A}}
\def\N{\mathbb{N}}
\def\T{\mathbb{T}}
\def\R{\mathbb{R}}
\def\Z{\mathbb{Z}}
\def\C{\mathbb{C}}
\def\O{\mathcal{O}}
\def\Sp{\mathcal{S}_+}
\def\Lap{\triangle}
\nwc{\lap}{\bigtriangleup}
\nwc{\rest}{\restriction}
\nwc{\Diff}{\operatorname{Diff}}
\nwc{\diam}{\operatorname{diam}}
\nwc{\Res}{\operatorname{Res}}
\nwc{\Spec}{\operatorname{Spec}}
\nwc{\Vol}{\operatorname{Vol}}
\nwc{\Op}{\operatorname{Op}}
\nwc{\supp}{\operatorname{supp}}
\nwc{\Span}{\operatorname{span}}

\nwc{\dia}{\varepsilon}
\nwc{\cut}{f}
\nwc{\qm}{u_\hbar}

\def\hto0{\xrightarrow{\hbar\to 0}}
\def\htoo{\stackrel{h\to 0}{\longrightarrow}}
\def\rto0{\xrightarrow{r\to 0}}
\def\rtoo{\stackrel{r\to 0}{\longrightarrow}}
\def\ntoinf{\xrightarrow{n\to +\infty}}

\providecommand{\abs}[1]{\lvert#1\rvert}
\providecommand{\norm}[1]{\lVert#1\rVert}
\providecommand{\set}[1]{\left\{#1\right\}}

\nwc{\la}{\langle}
\nwc{\ra}{\rangle}
\nwc{\lp}{\left(}
\nwc{\rp}{\right)}

\nwc{\bequ}{\begin{equation}}
\nwc{\ben}{\begin{equation*}}
\nwc{\bea}{\begin{eqnarray}}
\nwc{\bean}{\begin{eqnarray*}}
\nwc{\bit}{\begin{itemize}}
\nwc{\bver}{\begin{verbatim}}

%\nwc{\eal}{\end{align}}
\nwc{\eequ}{\end{equation}}
\nwc{\een}{\end{equation*}}
\nwc{\eea}{\end{eqnarray}}
\nwc{\eean}{\end{eqnarray*}}
\nwc{\eit}{\end{itemize}}
\nwc{\ever}{\end{verbatim}}

\newcommand{\defeq}{\stackrel{\rm{def}}{=}}

\title[Entropy of eigenfunctions]
{Entropy of eigenfunctions}

\author[N. Anantharaman]{Nalini Anantharaman}
\author[H. Koch]{Herbert Koch}
\author[S. Nonnenmacher]{St\'ephane Nonnenmacher}
\address{CMLS, \'Ecole Polytechnique, 91128 Palaiseau, France}
\email{nalini@math.polytechnique.fr}
\address{Mathematical Institute, University of Bonn,
Beringstra{\ss}e 1,D-53115 Bonn, Germany}
\email{koch@math.uni-bonn.de}
\address{Service de Physique Th\'eorique, 
CEA/DSM/PhT, Unit\'e de recherche associ\'ee au CNRS,
CEA/Saclay, 91191 Gif-sur-Yvette, France}
\email{snonnenmacher@cea.fr}

\begin{abstract}We study the high--energy limit for eigenfunctions of the laplacian, on a compact
negatively curved manifold. We review the recent result of Anantharaman--Nonnenmacher
\cite{AN07} giving a lower bound on the Kolmogorov--Sinai entropy of semiclassical measures.
The bound proved here improves the result of \cite{AN07} in the case of variable negative curvature.
\end{abstract}

\maketitle
\section{Motivations}
The theory of quantum chaos tries to understand how the chaotic behaviour of a classical
Hamiltonian system is reflected in its quantum counterpart. 
For instance, let $M$ be a compact Riemannian $C^\infty$ manifold, with negative 
sectional curvatures. The geodesic flow has the Anosov
property, which is considered as the ideal chaotic behaviour in the theory of dynamical systems. 
The corresponding quantum dynamics 
is the unitary flow generated by the Laplace-Beltrami operator on $L^2(M)$. One expects that the 
chaotic properties of the geodesic flow influence the spectral theory
of the Laplacian. The Random Matrix conjecture \cite{bohigas} asserts that the 
large eigenvalues should, after proper unfolding, statistically resemble those of a 
large random matrix, at least for a generic Anosov metric. 
The Quantum
Unique Ergodicity conjecture \cite{RudSar94} (see also \cite{berry77, voros77}) describes 
the corresponding eigenfunctions $\psi_k$: it claims that the probability measure
$|\psi_k(x)|^2 dx$ should approach (in the weak topology) the Riemannian volume,
when the eigenvalue tends to infinity. In fact a stronger property should hold for the 
{\em Wigner transform} $W_{\psi}$, a function on the cotangent bundle $T^* M$,
(the classical phase space)
which simultaneously describes the localization of the wave function $\psi$ in
position and momentum. 

We will adopt a semiclassical point of view, that is consider the
eigenstates of eigenvalue unity of the semiclassical Laplacian $-\hbar^2\lap$, thereby replacing
the high-energy limit by the
semiclassical limit $\hbar\to 0$.
We denote by $(\psi_k)_{k\in \N}$ an orthonormal basis
of $L^2(M)$ made of eigenfunctions of the Laplacian, and by $(-\frac1{\hbar_k^{2}})_{k\in \N}$
the corresponding eigenvalues: 
\begin{equation}\label{e:ef}
-\hbar_k^2\,\Lap \psi_k=\psi_k,\quad \text{with} \quad \hbar_{k+1}\leq \hbar_{k}\,.
\end{equation} 
We are interested in the high-energy eigenfunctions of $-\lap$, in other words the
semiclassical limit $\hbar_k\to 0$.

The Wigner distribution associated to an eigenfunction $\psi_k$ is defined by
$$
W_k(a)=\langle \Op_{\hbar_k}(a)\psi_k, \psi_k\rangle_{L^2(M)},\qquad a\in C_c^\infty(T^* M)\,.
$$
Here $\Op_{\hbar_k}$ is a quantization procedure, set at the scale (wavelength) $\hbar_k$, 
which associates to any smooth phase space function $a$ (with nice behaviour at infinity)
a bounded operator on $L^2(M)$. See for instance \cite{DS99} or \cite{EvZw06} 
for various quantizations
$\Op_\hbar$ on $\R^d$. On a manifold, one can use local
coordinates to define $\Op$ in a finite system of charts, then glue the objects
defined locally thanks to a smooth partition of unity \cite{CdV85}. 
For standard quantizations $\Op_{\hbar_k}$, the
Wigner distribution is of the form $W_k(x,\xi)\,dx\,d\xi$, where $W_k(x,\xi)$ is 
a smooth function on $T^*M$, 
called the Wigner transform of $\psi$. If $a$ is a function
on the manifold $M$, $\Op_\hbar(a)$ can be taken as the multiplication by $a$, and thus we have 
$W_k(a)= \int_M a(x) |\psi_k(x)|^2 dx$: the Wigner transform is thus
a {\em microlocal lift} of the density $|\psi_k(x)|^2$. 
Although the definition
of $W_k$ depends on a certain number of choices, like the choice of local coordinates,
or of the quantization procedure (Weyl, anti-Wick, ``right'' or ``left'' quantization...),
its asymptotic
behaviour when $\hbar_k\To 0$ does not. Accordingly, we call 
{\em semiclassical measures} the limit points
of the sequence $(W_k)_{k\in\IN}$, in the distribution topology.

In the semiclassical limit, ``quantum mechanics converges
to classical mechanics''. We will denote
$\abs{\cdot}_x$ the norm on $T^*_x M$ given by the metric.
The geodesic
flow $(g^t)_{t\in \R}$ is the Hamiltonian flow on $T^* M$ generated by the
Hamiltonian 
$H(x, \xi)=\frac{\abs{\xi}^2_x}2$. A quantization of this Hamiltonian is given
by the rescaled Laplacian $-\frac{\hbar^2\lap}2$, which
generates the unitary flow $(U_\hbar^t)= (\exp(it\hbar\frac\lap{2}))$ acting on 
$L^2(M)$. The semiclassical correspondence of the flows $(U_\hbar^t)$ and $(g^t)$ is expressed
through the Egorov Theorem~:
\begin{thm} \label{t:Egorov}
Let $a\in C_c^\infty(T^* M)$. Then, for any given $t$ in $\R$,
\begin{equation}\label{eg}\norm{U_\hbar^{-t}\Op_\hbar(a)U_\hbar^t-\Op_\hbar(a\circ g^t)}_{L^2(M)}=\cO(\hbar)\,,\qquad\hbar\to 0\,.
\end{equation}
\end{thm}
The constant implied in the remainder grows (often exponentially) with $t$, which represents
a notorious problem
when one wants to study the large time behaviour
of $(U_\hbar^t)$. Typically, the quantum-classical correspondence 
will break down for times $t$ of the order 
of the Ehrenfest time \eqref{e:Ehrenf}.

Using \eqref{eg} and other standard semiclassical arguments, one shows
the following~:
\begin{prop}\label{e:semiclass-measure} 
Any semiclassical measure is a probability measure carried on the energy
layer $\cE=H^{-1}(\frac12)$ (which coincides with the unit cotangent bundle $S^* M$). 
This measure is invariant under the geodesic flow.
\end{prop}
Let us call $\mathfrak{M}$ the set of $g^t$-invariant probability measures on $\cE$. This
set is convex and compact for the weak topology.
If the geodesic flow has the Anosov property --- for instance if $M$ has 
negative sectional curvature ---
that set is very large.
The geodesic flow has countably many periodic orbits, each of them
carrying an invariant probability measure. There are many other invariant measures, 
like the equilibrium states obtained by variational principles \cite{KatHas95}, among them
the Liouville measure $\mu_{\rm Liouv}$, and the measure of maximal entropy. 
Note that, for all these examples of measures,
the geodesic flow acts ergodically, meaning that these examples are extremal points in $\mathfrak{M}$.
Our aim is to determine, at least partially, the set  $\mathfrak{M}_{sc}$ formed by all possible 
semiclassical measures. 
 By its definition,
$\mathfrak{M}_{sc}$ is a closed subset of $\mathfrak{M}$, in the weak topology.

For manifolds such that the geodesic flow is ergodic with respect to the Liouville measure, 
it has been known for some time
that {\em almost all} eigenfunctions 
become equidistributed over $\cE$, in the semiclassical limit. This property 
is dubbed as Quantum Ergodicity~:
\begin{thm} \cite{Shni74,Zel87,CdV85}\label{t:QE}
Let $M$ be a compact Riemannian manifold, assume that the action
of the geodesic flow on $\cE=S^*M$ is ergodic with respect to the Liouville
measure. Let $(\psi_k)_{k\in\N}$ be an orthonormal basis of $L^2(M)$ consisting
of eigenfunctions of the Laplacian \eqref{e:ef}, and let $(W_k)$ be the associated
Wigner distributions on $T^*M$.

Then, there exists a subset $\cS\subset \N$ of density $1$, such that
\bequ\label{e:QE}
W_k\Lim \mu_{\rm Liouv},\qquad k\to\infty,\ k\in\cS.
\eequ
\end{thm}
The question of existence of ``exceptional'' subsequences of eigenstates 
with a different behaviour is still open. 
On a negatively curved manifold, the geodesic flow
satisfies the ergodicity assumption, and in fact much stronger
properties~: mixing, $K$--property, etc.
For such manifolds, it has been postulated in the Quantum Unique Ergodicity conjecture \cite{RudSar94}
that the full sequence of eigenstates becomes semiclassically equidistributed over $\cE$: 
one can take $\cS=\IN$ in the limit \eqref{e:QE}.
In other words, this conjecture states that
there exists a unique semiclassical measure, and $\mathfrak{M}_{sc}=\set{\mu_{\rm Liouv}}$.

So far the most precise results
on this question were obtained for manifolds $M$ with constant
negative curvature and {\em arithmetic} properties:
see Rudnick--Sarnak \cite{RudSar94},  Wolpert \cite{Wol01}. In that
very particular situation, there exists a countable commutative family
of self--adjoint operators commuting with the Laplacian~: the Hecke operators.
One may thus decide to restrict the attention to common bases of eigenfunctions, often
called ``arithmetic'' eigenstates, or Hecke eigenstates. A few years ago, 
Lindenstrauss \cite{Linden06} proved that any sequence of
arithmetic eigenstates become asymptotically equidistributed.
If there is some degeneracy in the spectrum of the Laplacian,
note that it could be possible that the Quantum Unique Ergodicity conjectured by Rudnick and Sarnak
holds for one orthonormal basis but not for another. 
On such arithmetic manifolds,
it is believed that the spectrum of the Laplacian has bounded multiplicity: if this is really the 
case, then the semiclassical equidistribution easily extends to any sequence of eigenstates.

Nevertheless, one may be less optimistic when extending the Quantum Unique Ergodicity conjecture
to more general systems.
One of the simplest example of a symplectic Anosov
dynamical system is given by linear hyperbolic automorphisms of the 2-torus, $e.g.$ Arnold's ``cat map''
$\begin{pmatrix}2 & 1 \\1 & 1 \\ \end{pmatrix}$. This system can be quantized
into a sequence of $N\times N$ unitary matrices --- the propagators, where $N\sim\hbar^{-1}$ \cite{HB80}.
The eigenstates of these matrices satisfy a Quantum Ergodicity theorem similar with Theorem~\ref{t:QE},
meaning that almost all eigenstates become equidistributed on the torus
in the semiclassical limit \cite{BouzDB96}. 
Besides, one can choose orthonormal eigenbases of the propagators, such that
the whole sequence of eigenstates is semiclassically equidistributed \cite{KurRud00}. 
Still, because the spectra of the propagators are highly degenerate, 
one can also construct
sequences of eigenstates with a different limit measure \cite{FNdB03}, for instance, 
a semiclassical measure consisting in two ergodic components:
half of it is the Liouville measure, while the other half is a Dirac peak on 
a single (unstable) periodic orbit. It was also shown
that this half-localization is {\em maximal} for this model \cite{FN04}~: 
a semiclassical measure cannot have more than half its mass carried by a countable union of periodic orbits.
The same type of half-localized eigenstates were constructed by two of the authors for another
solvable model, namely the ``Walsh quantization'' of the baker's map on the torus \cite{AN06}; for
that model, there exist ergodic semiclassical measures of purely fractal type (that is, without any
Liouville component).
Another type of semiclassical measure was recently obtained by Kelmer for quantized hyperbolic
automorphisms on higher-dimensional tori \cite{Kelmer05}: it
consists in the Lebesgue measure on some invariant co-isotropic 
subspace of the torus. 

For these Anosov models on tori, the construction of exceptional
eigenstates strongly uses nongeneric algebraic properties of the classical and quantized systems, and
cannot be generalized to nonlinear systems.

\section{Main result.}

In order to understand the set $\mathfrak{M}_{sc}$, we will attempt to compute 
the {\em Kolmogorov--Sinai} 
entropies of semiclassical measures. We work on a compact Riemannian manifold $M$ 
of arbitrary dimension, and
assume that the geodesic flow has the Anosov property. Actually, our method
can without doubt be adapted to more general Anosov Hamiltonian systems.

The Kolmogorov--Sinai entropy, also called
metric entropy, of a $(g^t)$-invariant probability measure $\mu$ is a nonnegative number $h_{KS}(\mu)$
that describes, in some sense, the complexity of a
$\mu$-typical orbit of the flow. The precise definition
will be given later, but for the moment let us just give a few facts. A measure carried on a closed geodesic
has vanishing entropy. In constant curvature, the entropy is maximal for the Liouville measure.
More generally, for any Anosov flow, 
the energy layer $\cE$ is foliated into unstable manifolds of the flow.
An upper bound on the entropy of an invariant probability measure is then provided 
by the Ruelle inequality:
\begin{equation}\label{Ruelle}
h_{KS}(\mu)\leq\left| \int_{\cE} \log J^u(\rho)d\mu(\rho)\right|\,.
\end{equation}
In this inequality, $J^u(\rho)$ is the {\em unstable Jacobian} of the flow at the point $\rho\in \cE$, 
defined as the Jacobian of the map $g^{-1}$ restricted to the unstable manifold at the
point $g^1 \rho$ (note that the average of $\log J^u$ over any invariant measure is negative).
The equality holds in \eqref{Ruelle} if and only if $\mu$ is the Liouville measure on $\cE$ \cite{LY85}.
If $M$ has dimension $d$ and has constant sectional curvature $-1$, the above inequality just reads
$h_{KS}(\mu)\leq d-1$. 

Finally, an important property of the metric entropy is that it
is an {\em affine} functional on $\mathfrak{M}$.
According to the Birkhoff ergodic theorem,
for any $\mu\in \mathfrak{M}$ and for $\mu$--almost every $\rho\in\cE$, 
the weak limit
$$
\mu^\rho=\lim_{|t|\To \infty}\frac1t \int_0^t \delta_{g^s \rho}ds
$$
exists, and is an ergodic probability measure. We can then write
$$
\mu=\int_{\cE} \mu^\rho d\mu(\rho),
$$
which realizes the ergodic decomposition of $\mu$.
The affineness of the KS entropy means that
$$
h_{KS}(\mu)=\int_{\cE} h_{KS}(\mu^\rho) d\mu(\rho).
$$
An obvious consequence is the fact that the range of $h_{KS}$ on $\mathfrak{M}$ is an
interval $[0,h_{max}]$.

\medskip

In the whole article, we consider a certain
subsequence of eigenstates $(\psi_{k_j})_{j\in\IN}$ of the Laplacian, such
that the corresponding sequence of Wigner distributions $(W_{k_j})$ converges
to a semiclassical measure $\mu$. In the following, the subsequence 
$(\psi_{k_j})_{j\in\IN}$
will simply be denoted by $(\psi_\hbar)_{\hbar\to 0}$, using the slightly abusive 
notation $\psi_\hbar=\psi_{\hbar_{k_j}}$ for the eigenstate $\psi_{k_j}$. 
Each eigenstate $\psi_\hbar$ thus satisfies
\begin{equation}\label{e:eigenstate}
(-\hbar^2\lap-1)\psi_\hbar=0\,.
\end{equation}
In \cite{An} the first author proved that the entropy of any $\mu\in\mathfrak{M}_{sc}$ 
is strictly positive. In \cite{AN07}, more explicit
lower bounds were obtained. The aim of this paper is to improve the lower bounds
of \cite{AN07} into the following
\begin{thm}\label{t:thetheorem} 
Let $\mu$ be a semiclassical measure
associated to the eigenfunctions of the Laplacian on $M$. Then its metric entropy satisfies
\begin{equation}\label{e:main1}
h_{KS}(\mu)\geq \left|\int_{\cE} \log J^u(\rho)d\mu(\rho) \right|- \frac{(d-1)}2\lambda_{\max}\,,
\end{equation}
where $d=\dim M$ and $\lambda_{\max}=\lim_{t\to\pm \infty}\frac{1}t
\log \sup_{\rho\in \cE} |dg^t_\rho|$ is the maximal expansion rate of the geodesic flow on $\cE$.

In particular, if $M$ has constant sectional curvature $-1$, we have
\begin{equation}\label{e:main2}
h_{KS}(\mu)\geq \frac{d-1}2.
\end{equation}
\end{thm}
In dimension $d$, we always have
$$ 
\left|\int_{\cE} \log J^u(\rho)d\mu(\rho) \right|\leq (d-1)\lambda_{\max}\,,
$$
so the above bound is an improvement over the one obtained in \cite{AN07},
\begin{equation}
h_{KS}(\mu)\geq \frac32\left|\int_{\cE} \log J^u(\rho)d\mu(\rho) \right|- (d-1)\lambda_{\max}\,.
\end{equation}
In the case of constant or little-varying curvature, 
the bound \eqref{e:main2} is much sharper than
the one proved in \cite{An}. On the other hand, 
if the curvature varies a lot (still being negative everywhere),
the right hand side of \eqref{e:main1} may actually be negative, 
in which case the bound is trivial.
We believe this ``problem'' to be a technical shortcoming of our method, 
and actually conjecture the following bound:
\begin{equation}\label{e:main3}
h_{KS}(\mu)\geq \frac12\left|\int_{\cE} \log J^u(\rho)d\mu(\rho) \right|\,.
\end{equation}
Extended to the case of the quantized torus automorphisms or the Walsh-quantized baker's map, 
this bound is saturated for the
half-localized semiclassical measures constructed in \cite{FNdB03}, as well as those obtained 
in \cite{Kelmer05, AN06}.
This bound allows certain ergodic components to be carried by closed geodesics, as long as
other components have positive entropy. This may be compared with the following
result obtained by Bourgain and Lindenstrauss in the case of arithmetic surfaces~:
\begin{thm}\cite{BLi03}
Let $M$ be a congruence arithmetic surface, 
and $(\psi_j)$ an orthonormal basis of eigenfunctions
for the Laplacian and the Hecke operators.

Let $\mu$ be a corresponding semiclassical measure, with ergodic decomposition
$\mu=\int_{\cE}\mu^\rho d\mu(\rho)$.  
Then, for $\mu$-almost all ergodic components we have $h_{KS}(\mu^\rho)\geq \frac19$.
\end{thm} 
As discussed above, the Liouville measure is the only one satisfying
$h_{KS}(\mu)= \left|\int_{\cE} \log J^u(\rho)\,d\mu(\rho) \right|$ \cite{LY85}, so 
the Quantum Unique Ergodicity would be proven in one could replace $1/2$ by $1$ on the right 
hand side of \eqref{e:main3}.
However, we believe that \eqref{e:main3}
is the optimal result that can be obtained without using much more precise information,
like for instance a sharp control on the spectral degeneracies, or fine information on
the lengths of closed geodesics. 

Indeed, in the above mentioned examples of Anosov systems where the
Quantum Unique Ergodicity conjecture is wrong and the
bound \eqref{e:main3} {\em sharp}, the quantum spectrum has very
high degeneracies, which could 
be responsible for the possibility to construct exceptional eigenstates.
Such high degeneracies are not expected in the case of the Laplacian 
on a negatively curved manifold. 
For the moment, however, there is no clear understanding
of the precise relation between spectral degeneracies and failure of Quantum Unique Ergodicity.

\subsection*{ Acknowledgements} 
N.A and S.N. were partially supported by the Agence Nationale de la Recherche, under
the grant ANR-05-JCJC-0107-01. They benefited from numerous discussions with Y.~Colin de Verdi\`ere
and M.~Zworski. S.N. is grateful to the Mathematical Department in Bonn for its hospitality in 
December 2006.

\section{Outline of the proof}

We start by recalling the definition and some properties of the metric entropy
associated with a probability measure on $T^*M$, invariant through
the geodesic flow. 
In \S\ref{s:class->qu} we extend the notion of entropy 
to the quantum framework. Our approach is semiclassical, so we want 
the classical and quantum entropies to be connected in some way when $\hbar\to 0$. 
The weights appearing in our quantum entropy are estimated in Thm.~\ref{t:mainestimate}, 
which was proven and used in \cite{An}. In \S\ref{s:connection} we also compare
our quantum entropy with several ``quantum dynamical entropies'' 
previously defined in the literature.
The proof of Thm.~\ref{t:thetheorem} actually starts in \S\ref{s:WEUP}, where we
present the algebraic tool allowing us to take advantage of 
our estimates \eqref{e:mainest} (or their optimized version given in Thm.~\ref{t:main}), 
namely an ``entropic uncertainty principle'' specific
of the quantum framework. From \S\ref{s:apply} on, we apply this
``principle'' to the quantum entropies appearing in our problem, and proceed to
prove Thm.~\ref{t:thetheorem}. Although the method is 
basically the same as in \cite{AN07}, several small modifications allow to finally
obtain the improved lower bound \eqref{e:main1}, and also simplify some intermediate proofs, 
as explained in Remark~\ref{r:improv}.

\subsection{Definition of the metric entropy}\label{s:KS-entropy}
In this paper we will meet several types of entropies, all of which are defined
using the function $\eta(s)=-s\log s$, for $s\in [0,1]$.
We start with the Kolmogorov-Sinai entropy of the geodesic flow with respect to an
invariant probability measure.

Let $\mu$ be a probability measure on the cotangent bundle $T^*M$.
Let  $\cP=(E_1,\ldots, E_K)$ be a finite measurable partition of $T^*M$~: 
$T^*M=\bigsqcup_{i=1}^K E_i$. We will denote the set of indices $\set{1,\ldots,K}=\setK$.
The Shannon entropy of $\mu$ with respect to the partition $\cP$ is defined as
$$
h_{\cP}(\mu)= - \sum_{k=1}^K \mu(E_k)\log\mu(E_k)=\sum_{k=1}^K \eta\big(\mu(E_k)\big).
$$
For any integer $n\geq 1$, we denote by $\cP^{\vee n}$ the partition
formed by the sets 
\bequ\label{e:hn}
E_{\bal}=E_{\alpha_0}\cap g^{-1}E_{\alpha_1}\ldots\cap g^{-n+1}E_{\alpha_{n-1}}\,,
\eequ
where $\bal=(\alpha_0,\ldots,\alpha_{n-1})$ can be
any sequence in $\setK^n$ (such a sequence is said to be of {\em length} $|\bal|=n$). 
The partition $\cP^{\vee n}$ is called the $n$-th refinement 
of the initial partition $\cP=\cP^{\vee 1}$.
The entropy of $\mu$ with respect to $\cP^{\vee n}$ is denoted by
\bequ \label{e:vee}
h_n(\mu, \cP)=h_{\cP^{\vee n}}(\mu)=\sum_{\bal\in\setK^n}
\eta\big(\mu(E_{\bal})\big)\,.
\eequ
If $\mu$ is $(g^t)$--invariant, it follows from
the convexity of the logarithm that
\begin{equation}\label{e:cl-subadd}
\forall n,m\geq 1,\qquad h_{n+m}(\mu, \cP)\leq h_n(\mu, \cP)+h_m(\mu, \cP),
\end{equation}
in other words the sequence $(h_n(\mu, \cP))_{n\in\N}$ is subadditive.
The entropy of $\mu$ with respect to the action of the geodesic flow and to 
the partition $\cP$ is defined by
\begin{equation}\label{e:def--ent}
h_{KS}(\mu, \cP)=\lim_{n\to +\infty}\frac{ h_n(\mu, \cP)}n=
\inf_{n\in\N}\frac{ h_n(\mu, \cP)}n.
\end{equation}
Each weight $\mu(E_{\bal})$ measures the $\mu$--probability
to visit successively $E_{\alpha_0},E_{\alpha_1},\ldots,E_{\alpha_{n-1}}$ at times 
$0,1,\ldots,n-1$ through the geodesic
flow.
Roughly speaking, the entropy measures the exponential decay of these probabilities
when $n$ gets large. It is easy to see that
$h_{KS}(\mu, \cP)\geq\beta$ if there exists $C$ such that 
$\mu(E_{\bal})\leq C\,e^{-\beta n}$, for all $n$ and all $\bal\in \bbleft 1,K\bbright^n$.

Finally, the Kolmogorov-Sinai entropy of $\mu$ with respect to the action of 
the geodesic flow is defined as
\begin{equation}\label{e:sup}
h_{KS}(\mu)=\sup_{\cP} h_{KS}(\mu,\cP),
\end{equation}
the supremum running over all finite measurable partitions $\cP$. 
The choice to consider the time $1$ of the geodesic flow in the definition \eqref{e:hn} may seem arbitrary,
but the entropy has a natural scaling property~: the entropy of $\mu$ with respect to the flow $(g^{at})$ is
$|a|$--times its entropy with respect to $(g^t)$.

Assume $\mu$ is carried on the energy layer $\cE$. Due to the Anosov property of 
the geodesic flow on $\cE$, it is known that the supremum \eqref{e:sup} is reached 
as soon as the diameter of the partition $\cP\cap \cE$
(that is, the maximum diameter of its elements $E_k\cap \cE$) is small enough. 
Furthermore, let us assume (without loss of generality) that the injectivity radius of 
$M$ is larger than $1$. Then, we may restrict our attention to partitions $\cP$ obtained by lifting on $\cE$
a partition of the manifold $M$, that is take $M=\bigsqcup_{k=1}^{K} M_k$ and then $E_k=T^*M_k$. 
In fact, if the diameter of $M_k$ in $M$ is of order $\dia$, then the diameter of 
the partition $\cP^{\vee 2}\cap\cE$ in $\cE$ is also of order $\dia$.
This special choice of our partition is not crucial, but it simplifies certain aspects of the analysis.

The existence of the limit in \eqref{e:def--ent}, and the fact that it coincides 
with the infimum, follow from a standard subadditivity argument. 
It has a crucial consequence~: if $(\mu_i)$ is a sequence of 
$(g^t)$--invariant probability measures on $T^*M$, weakly converging to a probability $\mu$, and 
if $\mu$ does not charge the boundary of the partition $\cP$, we have
$$
h_{KS}(\mu, \cP)\geq \limsup_i h_{KS}(\mu_i,\cP)\,.
$$
In particular, assume that for $i$ large enough,
the following estimates hold~:
\bequ\label{e:measure-decay}
\forall n\geq 1,\ \forall \bal\in\setK^n,\qquad
\mu_i(E_{\bal})\leq C_i \,\e^{-\beta n}\,,
\eequ
with $\beta$ independent of $i$. This implies for $i$ large enough
$h_{KS}(\mu_i,\cP)\geq\beta$,
and this estimate goes to the limit to yield
$h_{KS}(\mu)\geq\beta$.

\subsection{From classical to quantum dynamical entropy}\label{s:class->qu}
Since our semiclassical measure $\mu$ is defined as a limit
of Wigner distributions $W_\hbar$, a naive idea would be to estimate from below
the KS entropy of $W_\hbar$ and then take the limit $\hbar\to 0$. This
idea cannot work directly, because the Wigner transforms $W_\hbar$ are neither
positive, nor are they $(g^t)$--invariant. Therefore, one cannot
directly use the (formal) integrals $W_\hbar(E_{\bal})=\int_{E_{\bal}}W_\hbar(x,\xi)\,dx\,d\xi$
to compute the entropy of the semiclassical measure.

Instead, the method initiated by the first author in \cite{An}
is based on the following remarks. Each integral $W_\hbar(E_{\bal})$ can also be written as 
$W_\hbar(\bbbone_{E_{\bal}})=\int_{T^*M} W_\hbar\,\bbbone_{E_{\bal}}$, 
where $\bbbone_{E_{\bal}}$ is the characteristic function
on the set $E_{\bal}$, that is
\bequ\label{e:product-charact}
\bbbone_{E_{\bal}}=(\bbbone_{E_{\alpha_{n-1}}}\circ g^{n-1})\times \ldots\times
(\bbbone_{E_{\alpha_1}}\circ g)\times \bbbone_{E_{\alpha_0}}\,.
\eequ
Remember we took $E_k=T^* M_k$, where the $M_k$ form a partition of $M$.

From the definition of the Wigner distribution, this integral corresponds formally to the overlap
$\la \psi_\hbar, \Op_\hbar(\bbbone_{E_{\bal}})\psi_\hbar\ra$. Yet, the characteristic
functions $\bbbone_{E_{\bal}}$ have sharp discontinuities, so their 
quantizations cannot be incorporated in a nice pseudodifferential calculus. Besides, 
the set $E_{\bal}$ is not
compactly supported, and shrinks in the unstable direction when
$n=\abs{\bal}\To+ \infty$, so that the operator $\Op_\hbar(\bbbone_{E_{\bal}})$ is very problematic.

We also note that an overlap of the form 
$\la \psi_\hbar, \Op_\hbar(\bbbone_{E_{\bal}})\psi_\hbar\ra$ is a {\it hybrid} expression: 
this is a {\em quantum} matrix element of an
operator defined in terms of the {\em classical}
evolution \eqref{e:product-charact}. From the point of view of quantum mechanics, it is more natural 
to consider, instead, the operator obtained as the product of Heisenberg-evolved quantized functions, 
namely
\bequ\label{e:Qproduct-charact}
(U_\hbar^{-n+1}P_{\alpha_{n-1}}U_\hbar^{n-1})\, 
(U_\hbar^{-n+2}P_{\alpha_{n-2}}U_\hbar^{n-2})\,
\cdots (U_\hbar^{-1}P_{\alpha_1}U_\hbar)\,P_{\alpha_0}\,.
\eequ
Here we used the shorthand notation $P_k= \bbbone_{M_k}$, $k\in\setK$ (multiplication
operators). To remedy the fact that the functions $\bbbone_{M_k}$ are not smooth, which
would prevent us from using a semiclassical calculus, we
apply a convolution kernel to smooth them, obtain functions $\bbbone_{M_k}^{sm}\in C^\infty(M)$,
and consider $P_k\defeq \bbbone_{M_k}^{sm}$ 
(we can do this keeping the property $\sum_{k=1}^K\bbbone_{M_k}^{sm}=1$).

In the following, we will use 
the notation $A(t)\defeq U_\hbar^{-t}\,A\,U_\hbar^t$ for the 
Heisenberg evolution of the operator $A$ though the Schr\"odinger 
flow $U_\hbar^t= \exp(-it\hbar\frac\lap{2})$. The norm $\norm{\bullet}$ will denote
either the Hilbert norm on $L^2(M)$, or the corresponding operator norm.
The subsequent ``purely quantum'' norms were estimated in \cite[Thm.~1.3.3]{An}: 
\begin{thm} \label{t:mainestimate}{\em (The main estimate \cite{An})}
Set as above $P_k\defeq \bbbone_{M_k}^{sm}$.
For every ${\cK}>0$, there exists $\hbar_{\cK}>0$ such that,
uniformly for all $\hbar<\hbar_{\cK}$, for all $n\leq {\cK}|\log \hbar|$,
for all $(\alpha_0,\ldots,\alpha_{n-1})\in \setK^n$,
\bequ\label{e:mainest}
\norm{P_{\alpha_{n-1}}(n-1)\,P_{\alpha_{n-2}}(n-2)\cdots 
P_{\alpha_{0}}\,\psi_\hbar}
\leq 2(2\pi \hbar)^{-d/2}\,\e^{-\frac{\Lambda}2n}(1+\cO(\dia))^n.
\eequ
\end{thm}
The exponent $\Lambda$ is given by the ``smallest expansion rate'':
$$
\Lambda=-\sup_{\nu\in\mathfrak{M}} 
\int \log J^u(\rho)d\nu(\rho)=\inf_\gamma \sum_{i=1}^{d-1}\lambda_i^+(\gamma).
$$
The infimum on the right hand side runs over the set of closed orbits on $\cE$, 
and the $\lambda_i^+$ denote the positive Lyapunov
exponents along the orbit, that is the logarithms of the expanding eigenvalues of the Poincar\'e map,
divided by the period of the orbit. The parameter $\dia>0$ is an upper bound 
on the diameters of the supports of the functions
$\bbbone_{M_k}^{sm}$ in $M$.

From now on we will
call the product operator
\bequ\label{e:P_eps}
P_{\bal}=P_{\alpha_{n-1}}(n-1)\,P_{\alpha_{n-2}}(n-2)\cdots P_{\alpha_{0}},\qquad \bal\in\setK^n\,.
\eequ
To prove the above estimate, 
one actually controls the operator norm
\bequ\label{e:ineg1}
\norm{P_{\bal}\Op_\hbar(\chi)}\leq 2(2\pi \hbar)^{-d/2}\,
\e^{-\frac{\Lambda}2n}(1+\cO(\dia))^n\,,
\eequ
where $\chi\in C^\infty_c(\cE^\vareps)$ is an energy cutoff such that $\chi=1$ near $\cE$, supported
inside a neighbourhood $\cE^\vareps=H^{-1}([\frac12-\dia, \frac12+\dia])$ of $\cE$.

In quantum mechanics, the matrix element $\la\psi_\hbar,P_{\bal}\psi_\hbar\ra$ looks like
the ``probability'', for a particle in the state $\psi_\hbar$, to visit successively 
the phase space regions $E_{\alpha_0},E_{\alpha_1},\ldots,E_{\alpha_{n-1}}$ at times $0,1,\ldots,n-1$ 
of the Schr\"odinger flow. 
Theorem~\ref{t:mainestimate} implies that this ``probability'' 
decays exponentially fast with $n$, with rate $\frac\Lambda{2}$, but this decay only starts
around the time  
\bequ\label{e:Ehrenf0}
n_1\defeq\frac{d|\log\hbar|}{\Lambda}\,,
\eequ
which is a kind of ``Ehrenfest time'' (see \eqref{e:Ehrenf} for another definition of
Ehrenfest time). 

Yet, because the matrix elements
$\la\psi_\hbar,P_{\bal}\psi_\hbar\ra$ are not real in general, 
they can hardly be used to define a ``quantum measure''.
Another possibility to define the probability for the 
particle to visit the sets $E_{\alpha_k}$ at times $k$, is to take the squares of the norms appearing in 
\eqref{e:mainest}:
\bequ\label{e:qu-weights}
\norm{P_{\bal}\,\psi_\hbar}^2=\norm{P_{\alpha_{n-1}}(n-1)\,P_{\alpha_{n-2}}(n-2)\cdots 
P_{\alpha_{0}}\psi_\hbar}^2\,.
\eequ
Now we require the smoothed characteristic functions $\bbbone_{M_i}^{sm}$
to satisfy the identity
\bequ\label{e:sm-partition2}
\sum_{k=1}^K \big(\bbbone_{M_k}^{sm}(x)\big)^2=1\quad\text{ for any point } x\in M\,.
\eequ
We denote by $\cP_{sm}$ the smooth partition of $M$ made by the functions 
$\big((\bbbone_{M_k}^{sm})^2\big)_{k=1}^K$.
The corresponding set of multiplication operators $(P_k)_{k=1}^K\defeq\cP_q$ 
forms a ``quantum partition of unity''~:
\bequ\label{e:qu-part0}
\sum_{k=1}^K P_k^2=Id_{L^2}\,.
\eequ
For any $n\geq 1$, we refine the quantum partition $\cP_q$ into $(P_{\bal})_{|\bal|}$, as in \eqref{e:P_eps}.
The weights \eqref{e:qu-weights} exactly add up to unity, so
it makes sense to consider the entropy
\bequ\label{e:qu-entropy0}
h_n(\psi_\hbar,\cP_q) \defeq
\sum_{\bal\in\setK^n}\eta\big( \norm{P_{\bal}\,\psi_\hbar}^2 \big)\,.
\eequ

\subsubsection{Connection with other quantum entropies}\label{s:connection}

This entropy appears to be a particular case of the ``general quantum entropies'' 
described by S{\l}omczy{\'n}ski and \.{Z}yczkowski \cite{SloZy94},
who already had in mind applications to quantum chaos. In their terminology,
a family of bounded operators $\pi=(\pi_k)_{k=1}^{\cN}$ on a Hilbert space $\cH$
satisfying 
\begin{equation}\label{e:unity}
\sum_{k=1}^{\cN}\pi_k^{*}\,\pi_k=Id_{\cH}
\end{equation}
provides an ``instrument'' which, to each index $k\in \setN$, associates the following
map on density matrices:
$$
\rho\mapsto \cI(k)\rho= \pi_k\,\rho\,\pi_k^*\,,\quad\text{a nonnegative operator with }
\tr(\cI(k)\rho)\leq 1\,.
$$
From a unitary propagator $U$ and its adjoint action $\cU\rho=U\rho U^{-1}$, they propose
to construct the refined instrument
$$
\cI(\bal)\rho\defeq \cI(\alpha_{n-1})\circ \cdots \cU\circ\cI(\alpha_1)\circ\cU\circ\cI(\alpha_0)\rho=
U^{-n+1}\,\pi_{\bal}\,\rho\,\pi_{\bal}^*\,U^{n-1}\,,\qquad \bal\in\setN^n\,,
$$
where we used \eqref{e:P_eps} to refine the operators $\pi_k$ into $\pi_{\bal}$. 
We obtain the probability weights
\bequ\label{e:weights-diag}
\tr (\cI(\bal)\rho)=\tr(\pi_{\bal}\rho\pi_{\bal}^*)\,,\qquad \bal\in\setN^n.
\eequ
For any $U$-invariant density $\rho$, these weights provide an entropy
\bequ\label{e:SZ}
h_n(\rho,\cI)=\sum_{\bal\in \setN^n} \eta\big(\tr(\cI(\bal)\rho)\big)\,.
\eequ
One easily checks that our quantum partition $\cP_q=(P_k)_{k=1}^K$ satisfies \eqref{e:unity},
and that if one takes $\rho=|\psi_\hbar\ra\la\psi_\hbar|$ the weights 
$\tr (\cI(\bal)\rho)$ exactly correspond to our weights
$\norm{P_{\bal}\psi}^2$. Hence,
the entropy \eqref{e:SZ} coincides with \eqref{e:qu-entropy0}.

Around the same time, Alicki and Fannes \cite{AlFa94} used the same quantum
partition \eqref{e:unity} (which they called ``finite operational partitions of unity'') 
to define a different type of
entropy, now called the ``Alicki-Fannes entropy''
(the definition extends to general $C^*$-dynamical systems).
For each $n\geq 1$ they extend the weights \eqref{e:weights-diag} to ``off-diagonal entries''
to form a $\cN^n\times \cN^n$ density matrix $\rho_n$:
\bequ\label{e:off-diag}
[\rho_n]_{\bal',\bal}=\tr (\pi_{\bal'}\,\rho\,\pi_{\bal}^*),\qquad \bal,\bal'\in \bbleft 1,\cN\bbright^n\,.
\eequ
The AF entropy of the system $(\cU,\rho)$ is then defined as follows: take
the Von Neumann entropy of these density matrices, $h^{AF}_n(\rho,\pi)=\tr\,\eta(\rho_n)$, 
then take $\limsup_{n\to\infty}\frac1n h^{AF}_n(\rho,\pi)$ and finally take the supremum
over all possible finite operational partitions of unity $\pi$. 

We mention that traces of the form \eqref{e:off-diag} also appear in  
the ``quantum histories'' approach to quantum mechanics (see e.g. \cite{Gri84}, and 
\cite[Appendix~D]{SloZy94} for references).

\subsubsection{Naive treatment of the entropy $h_n(\psi_\hbar,\cP_q)$}\label{s:naive}
For fixed $|\bal|>0$, the Egorov theorem shows that $\norm{P_{\bal}\psi_\hbar}^2$
converges to the classical weight $\mu\big((\bbbone_{M_{\bal}}^{sm})^2\big)$ when $\hbar\to 0$, 
so for fixed $n>0$ the entropy $h_n(\psi_\hbar,\cP_q)$ converges to $h_n(\mu,\cP_{sm})$, defined
as in \eqref{e:vee}, the characteristic functions $\bbbone_{M_k}$ being replaced by their
smoothed versions $(\bbbone^{sm}_{M_k})^2$.
On the other hand, from the estimate \eqref{e:ineg1} the entropies $h_n(\psi_\hbar,\cP_q)$
satisfy, for $\hbar$ small enough,
\bequ\label{e:lower-hn}
h_n(\psi_\hbar,\cP_q)\geq n\big(\Lambda+\cO(\dia)\big)- d |\log\hbar| +\cO(1)\,, 
\eequ
for any time $n\leq \cK|\log\hbar|$. For large times $n\approx \cK|\log\hbar|$, this provides
a lower bound 
$$
\frac1n\, h_n(\psi_\hbar,\cP_q)\geq \big(\Lambda+\cO(\dia)\big)- \frac{d}{\cK}+\cO(1/|\log\hbar|)\,,
$$
which looks very promising since $\cK$ can be taken arbitrary large: 
we could be tempted to take the semiclassical limit, and 
deduce a lower bound $h_{KS}(\mu)\geq \Lambda$.

Unfortunately, this does not work, because in the range $\set{n>n_1}$ where the 
estimate \eqref{e:lower-hn} is useful,
the Egorov theorem breaks down, the weights \eqref{e:qu-weights} do not approximate the classical weights 
$\mu\big((\bbbone_{M_{\bal}}^{sm})^2\big)$, and there is no relationship between  
$h_n(\psi,\cP_q)$ and the classical entropies $h_n(\mu,\cP_{sm})$. 

This breakdown of the quantum-classical correspondence around the
Ehrenfest time is ubiquitous for chaotic dynamics. It has been observed before when 
studying the connection between the Alicki-Fannes entropy for the quantized torus automorphisms
and the KS entropy of the classical dynamics \cite{BCDCFVP03}: the quantum entropies 
$h^{AF}_n(\psi_\hbar,\cP_q)$
follow the classical $h_n(\mu,\cP_{sm})$ until the Ehrenfest time (and therefore grow linearly
with $n$), after which they ``saturate'', to produce a vanishing entropy
$\limsup_{n\to\infty}\frac1n h^{AF}_n(\psi_\hbar,\cP_q)$.

To prove the Theorem \ref{t:thetheorem}, we will still use the estimates \eqref{e:ineg1},
but in a more subtle way, namely by referring to an {\em entropic uncertainty principle}.

\subsection{Entropic uncertainty principle}\label{s:WEUP}

The theorem below is an adaptation of the entropic uncertainty principle conjectured by Deutsch
and Kraus \cite{Deu83,Kraus87} and proved by Massen and Uffink \cite{MaaUff}. These authors
were investigating the theory of
measurement in quantum mechanics.
Roughly speaking, this result states that if a unitary matrix has ``small'' entries, then
any of its eigenvectors must have a ``large'' Shannon entropy.

Let $(\cH, \la.,.\ra)$ be a complex Hilbert space, and denote $\norm{\psi}=\sqrt{\la\psi,\psi\ra}$ 
the associated norm.
Consider a quantum partition of unity $(\pi_k)_{k=1}^{\cN}$ on $\cH$ as in \eqref{e:unity}.
If $\norm{\psi}=1$, we define the entropy of $\psi$
with respect to the partition $\pi$ as in \eqref{e:qu-entropy0}, namely
$h_\pi(\psi)=\sum_{k=1}^{\cN}\eta\big(\norm{\pi_k\,\psi}^2\big)$.
We extend this definition by introducing the notion of {\em pressure}, associated to a family 
$\bv=(v_k)_{k=1,\ldots,\cN}$ of positive real numbers: 
the pressure is defined by
$$
p_{\pi,\bv}(\psi)\defeq \sum_{k=1}^\cN \eta\big(\norm{\pi_k\,\psi}^2\big) - 
\sum_{k=1}^\cN \norm{\pi_k\,\psi}^2\log v_k^2.
$$
In Theorem \ref{t:WEUP}, we actually need two partitions of unity
$ (\pi_k)_{k=1}^{\cN}$ and $(\tau_j)_{j=1}^{\cM}$,
and two families of weights $\bv=(v_k)_{k=1}^\cN$,
$\bw=(w_j)_{j=1}^\cM$, and consider the corresponding pressures $p_{\pi,\bv}(\psi),
p_{\tau,\bw}(\psi)$.
Besides the appearance of the weights
$\bv,\ \bw$, we bring another modification to the statement in \cite{MaaUff}
by introducing an auxiliary operator $\cO$.
\begin{thm}\cite[Thm.~6.5]{AN07}\label{t:WEUP}
Let $\cO$ be a bounded operator and $\cU$ be an isometry on $\cH$.\\ 
Define $c_\cO^{(\bv, \bw)}(\cU)\defi\sup_{j,k}  w_j\,v_k\, \norm{\tau_j\,\cU\,\pi_k^*\, \cO}$,
and $V=\max_k v_k$, $W=\max_j w_j$.
 
Then, for any $\eps\geq 0$, for any normalized $\psi\in\cH$ satisfying
\bequ\label{e:locali}
\forall k=1,\ldots,\cN,\qquad \norm {(Id-\cO)\,\pi_k\,\psi}\leq \eps\,,
\eequ
the pressures $p_{\tau,\bw}\big(\cU\psi \big),\, p_{\pi,\bv}\big(\psi\big)$ satisfy
$$
p_{\tau,\bw}\big(\cU\,\psi \big) + p_{\pi,\bv}\big(\psi\big)
\geq - 2 \log \big(c_\cO^{(\bv, \bw)}(\cU)+\cN\,V\,W\,\eps\big)\,.
$$
\end{thm}
\begin{ex}The original result of \cite{MaaUff} corresponds to the case where $\cH=\IC^\cN$,
$\cO=Id$, $\eps=0$, $\cN=\cM$, $v_k=w_j=1$,
and the operators $\pi_k=\tau_k$ are the orthogonal projectors on some orthonormal basis 
$(e_k)_{k=1}^{\cN}$ of $\cH$. In this case, the theorem asserts that
$$
h_\pi(\cU\,\psi)+h_\pi(\psi) \geq -2\log c(\cU)
$$
where $c(\cU)=\sup_{j, k} |\langle e_k, \cU e_j\rangle|$ is the supremum of all matrix elements of 
$\cU$ in the orthonormal basis $(e_k)$.
As a special case, one gets
$h_\pi(\psi)\geq -\log c(\cU)$ if $\psi$ is an eigenfunction of $\cU$.
\end{ex}

\subsection{Applying the entropic uncertainty principle to the Laplacian eigenstates}\label{s:apply}
In this section we explain how to use Theorem~\ref{t:WEUP} in order to obtain
nontrivial information on the quantum entropies \eqref{e:qu-entropy0} and then $h_{KS}(\mu)$.
For this we need to define the data to input in the theorem.
Except the Hilbert space $\cH=L^2(M)$, all other data depend on the semiclassical parameter $\hbar$:
the quantum partition $\pi$, the operator $\cO$, the positive real number $\eps$, the
weights $(v_j)$, $(w_k)$ and the unitary operator $\mathcal{U}$.
 
As explained in section~\ref{s:class->qu}, we partition $M$ into
$M=\sqcup_{k=1}^K M_k$,
consider open sets $\Omega_k\supset\!\!\supset M_k$ (which we assume to have diameters $\leq\dia$), 
and consider smoothed characteristic functions $\bbbone_{M_k}^{sm}$ supported respectively inside 
$\Omega_k$, and satisfying the identity \eqref{e:sm-partition2}. The associated multiplication
operators on $\cH$ are form a quantum partition $(P_k)_{k=1}^K$, which we had called $\cP_q$.
To alleviate notations, we will drop the subscript $q$.

From \eqref{e:qu-part0}, and using the unitarity of $U_\hbar$, one realizes that 
for any $n\geq 1$, the families of operators $\cP^{\vee n}=(P_{\bal}^*)_{|\bal|=n}$ and 
$\cT^{\vee n}=(P_{\bal})_{|\bal|=n}$ (see \eqref{e:P_eps})
make up two quantum partitions of unity as in \eqref{e:unity}, of cardinal $K^n$.

\subsubsection{Sharp energy localization}\label{s:energy}
In the estimate \eqref{e:ineg1},
we introduced an energy cutoff $\chi$ on a finite energy strip $\cE^{\dia}$, with $\chi\equiv 1$
near $\cE$. This cutoff
does not appear in the estimate \eqref{e:mainest}, because, when applied to the
eigenstate $\psi_\hbar$, the operator $\Op_\hbar(\chi)$ essentially acts like the identity.

The estimate \eqref{e:ineg1} will actually not suffice to prove Theorem~\ref{t:thetheorem}. We will need
to optimize it by replacing $\chi$ in \eqref{e:ineg1} with
a ``sharp'' energy cutoff. 
For some fixed (small) $\delta\in (0, 1)$, we consider a smooth function 
$\chi_\delta\in C^\infty(\IR;[0,1])$,
with $\chi_\delta(t)= 1$ for $|t|\leq \e^{-\delta/2}$ and $\chi_\delta(t)=0$ for $|t|\geq 1$. 
Then, we rescale that function to obtain the following family of
$\hbar$-dependent cutoffs near $\cE$:
\begin{equation}\label{e:chi_n}
\forall \hbar\in(0,1),\ \forall n\in\IN,\ 
\forall \rho\in T^*M,\qquad\chi^{(n)}(\rho;\hbar)\defeq 
\chi_\delta\big(\e^{-n\delta}\,\hbar^{-1+\delta}(H(\rho)-1/2)\big)\,.
\end{equation}
The cutoff $\chi^{(n)}$ is supported in a tubular neighbourhood of $\cE$ of width 
$ 2\hbar^{1-\delta}\,\e^{n\delta}$. We will always assume that this width is $<<\hbar^{1/2}$
in the semiclassical limit, which is the case if we ensure that $n\leq C_\del|\log\hbar|$
for some $0<C_\del<(2\delta)^{-1}-1$.
In spite of their singular behaviour,
these cutoffs can be quantized into pseudodifferential operators
$\Op(\chi^{(n)})$ described in \cite{AN07} (the quantization
uses a pseudodifferential calculus adapted to the energy layer $\cE$, drawn from \cite{SZ99}). 
The eigenstate $\psi_\hbar$ is indeed very localized near $\cE$, since it 
satisfies
\bequ\label{e:local}
\norm{\big(\Op(\chi^{(0)})-1\big)\psi_\hbar}=\cO(\hbar^\infty)\,\norm{\psi_\hbar}\,.
\end{equation} 
In the rest of the paper, we also fix a small $\delta'>0$, and call
``Ehrenfest time'' the $\hbar$-dependent integer
\begin{equation}\label{e:Ehrenf}
n_E(\hbar)\defeq\big\lfloor\frac{(1-\delta')|\log \hbar|}{\lambda_{\max}}\big\rfloor\,.
\end{equation}
Notice the resemblance with the time $n_1$ defined in \eqref{e:Ehrenf0}.
The significance of this time scale will be discussed in \S\ref{s:subadd}.

The following proposition states that the operators $(P^*_{\bal})_{|\bal|=n_E}$,
almost preserve the energy localization of $\psi_\hbar$~:
\begin{prop}
For any $L>0$, there exists $\hbar_L$ such that, for any $\hbar\leq \hbar_L$, 
the Laplacian eigenstate satisfies
\bequ\label{e:psi-cone}
 \forall \bal, |\bal|=n_E,\qquad
\norm{\big(\Op(\chi^{(n_E)})-Id \big) P^*_{\bal}\,\psi_\hbar}\leq \hbar^L\norm{\psi_\hbar}\,. 
\eequ
\end{prop}
We recognize here a condition of the form \eqref{e:locali}.

\subsubsection{Applying Theorem~\ref{t:WEUP}: Step 1}\label{s:mainapp}
We now precise some of 
the data we will use in the entropic uncertainty principle, Theorem~\ref{t:WEUP}. As opposed to
the choice made in \cite{AN07}, we will use two different partitions $\pi,\ \tau$.
\begin{itemize}
\item the quantum partitions $\pi$ and $\tau$ are given respectively
by the families of operators $\pi=\cP^{\vee n_E}=(P_{\bal}^*)_{|\bal|=n_E}$, 
$\tau=\cT^{\vee n_E}=(P_{\bal})_{|\bal|=n_E}$. Notice that these partitions only differ by
the ordering of the operators $P_{\alpha_i}(i)$ inside the products.
In the semiclassical limit, these partitions 
have cardinality $\cN=K^{n_E}\asymp \hbar^{-K_0}$ for some fixed $K_0>0$.
\item the isometry will be the propagator at the Ehrenfest time, $\cU=U_\hbar^{n_E}$. 
\item the auxiliarly operator is given as $\cO=\Op(\chi^{(n_E)})$, and the error 
$\eps=\hbar^L$, where $L$ will be chosen very large (see \S \ref{s:weights}).
\item the weights $v_{\bal},\ w_{\bal}$ will be selected in \S\ref{s:weights}. They will be 
semiclassically tempered, meaning that there exists $K_1>0$ such that, for $\hbar$ small enough, 
all $v_{\bal},\ w_{\bal}$ are contained in the interval $[1,\hbar^{-K_1}]$.
\end{itemize}
The entropy and pressures associated with a state $\psi\in\cH$ are given by
\begin{align}\label{e:entropy}
h_{\pi}(\psi)&=\sum_{|\bal|=n_E}\eta\big(\norm{P^*_{\bal}\,\psi}^2\big),\\
p_{\pi, \bv}(\psi)&= h_{\pi}(\psi)
-2\sum_{|\bal|=n_E}\norm{P^*_{\bal}\,\psi}^2\log v_{\bal}.\label{e:Jpressures}
\end{align}
With respect to the second partition, we have
\begin{align}
h_{\tau}(\psi)&=\sum_{|\bal|=n_E}\eta\big(\norm{P_{\bal}\,\psi}^2\big),\\
p_{\tau, \bw}(\psi)&= h_{\tau}(\psi)
-2\sum_{|\bal|=n_E}\norm{P_{\bal}\,\psi}^2\log w_{\bal}.
\end{align}
We notice that the entropy $h_{\tau}(\psi)$ exactly corresponds to the formula \eqref{e:qu-entropy0}, 
while $h_{\pi}(\psi)$ is built from the norms
$$
\norm{P_{\bal}^*\,\psi}^2=
\norm{P_{\alpha_0} P_{\alpha_1}(1) \cdots P_{\alpha_{n-1}}(n-1)\,\psi}^2\,.
$$
If $\psi$ is an eigenfunction of $U_\hbar$, the above norm can be obtained from \eqref{e:qu-weights}
by exchanging $U_\hbar$ with $U_\hbar^{-1}$, and replacing the sequence 
$\bal=(\alpha_0,\ldots,\alpha_{n-1})$
by $\bar{\bal}\defeq (\alpha_{n-1},\ldots,\alpha_0)$. 
So the entropies $h_{\pi}(\psi)$ and $h_{\tau}(\psi)$
are mapped to one another through the time reversal $U_\hbar \to U_\hbar^{-1}$.

With these data,
we draw from Theorem~\ref{t:WEUP} the following
\begin{cor}\label{c:WEUP} 
For $\hbar>0$ small enough consider the data $\pi$, $\tau$, $\cU$, $\cO$ as defined above. Let
\bequ\label{e:c_C}
c_{\cO}^{\bv, \bw}(\cU)\defeq
\max_{|\bal|=|\bal'|=n_E} \Big( w_{\bal'}\,v_{\bal}\,
\norm{P_{\bal'}\,U_\hbar^{n_E}\,P_{\bal} \Op(\chi^{(n_E)})}\Big)\,.
\end{equation}
Then
for any normalized state $\phi$ satisfying \eqref{e:psi-cone},
$$
p_{\tau, \bw} (U_\hbar^{n_E}\,\phi) + p_{\pi, \bv}(\phi) \geq 
-2\log\left(c_{\cO}^{\bv, \bw}(\cU)+h^{L-K_0-2K_1}\right)\,.
$$
\end{cor}
From \eqref{e:psi-cone}, we see that the above corollary applies to the eigenstate $\psi_\hbar$ if
$\hbar$ is small enough.

The reason to take the same value $n_E$ for the refined partitions $\cP^{\vee n_E}$, 
$\cT^{\vee n_E}$ and
the propagator $U_\hbar^{n_E}$ is the following~: the products appearing in 
$c_{\cO}^{\bv, \bw}(\cU)$ can be rewritten (with $U\equiv U_\hbar$):
$$
P_{\bal'}\,U^{n_E}\,P_{\bal}
= U^{-n_E+1}P_{\alpha'_{n_E-1}}U\cdots UP_{\alpha'_0}UP_{\alpha_{n_E-1}}U\cdots UP_{\alpha_0}
= U^{n_E}\,P_{\bal\bal'}\,.
$$
Thus, the estimate \eqref{e:ineg1} with $n=2n_E$ already provides an upper bound
for the norms appearing in \eqref{e:c_C} ---
the replacement of $\chi$ by the sharp cutoff $\chi^{(n_E)}$ does not harm the estimate. 

To prove Theorem~\ref{t:thetheorem}, 
we actually need to improve the estimate \eqref{e:ineg1}, as was done in \cite{AN07}, see 
Theorem~\ref{t:main} below. This improvement is done at two levels: we will use the fact that the
cutoffs $\chi^{(n_E)}$ are sharper than $\chi$, and also the fact that the expansion rate of the geodesic flow
(which governs the upper bound in \eqref{e:ineg1}) is not uniform, but 
depends on the sequence ${\bal}$.

Our choice for 
the weights $v_{\bal}$, $w_{\bal}$ will then be guided by
the $\bal$-dependent upper bounds given in Theorem~\ref{t:main}. To state that theorem, we
introduce some notations.

\subsubsection{Coarse-grained unstable Jacobian}\label{s:unstable-Jac}
We recall that, for any energy $\lambda>0$, the geodesic flow $g^t$ on
the energy layer $\cE(\lambda)=H^{-1}(\lambda)\subset T^*M$ is Anosov, so that
the tangent space $T_\rho\cE(\lambda)$ at each $\rho\in T^*M$, $H(\rho)>0$ splits into
$$
T_\rho\cE(\lambda)=E^u(\rho)\oplus E^s(\rho) \oplus \IR\,X_H(\rho)\,
$$
where $E^u$ (resp. $E^s$) is the unstable (resp. stable) subspace.
The unstable Jacobian $J^u(\rho)$ is defined by $J^u(\rho)=\det\big(dg^{-1}_{|E^u(g^1\rho)}\big)$
(the unstable spaces at $\rho$ and $g^1\rho$ are equipped with the induced Riemannian metric). 

This Jacobian can be ``discretized'' as follows in the energy strip
$\cE^{\dia}\supset \cE$. 
For any pair of indices $(\alpha_0,\alpha_1)\in \setK^2$, we define
\begin{equation}\label{e:coarse-Jac}
J^u_1(\alpha_0,\alpha_1)\defi \sup\set{J^u(\rho)\ :\ \rho\in T^*\Omega_{\alpha_0}\cap
\cE^{\dia},\ g^1 \rho\in T^*\Omega_{\alpha_1}}
\end{equation}
if the set on the right hand side is not empty, and $J^u_1(\alpha_0,\alpha_1)=e^{-R}$ otherwise,
where $R>0$ is a fixed large number.
For any sequence of symbols $\bal$ of length $n$, we define
\begin{equation}\label{e:multi}
J^u_n(\bal)\defi J^u_1(\alpha_0,\alpha_1)\cdots J^u_1(\alpha_{n-2},\alpha_{n-1})\,.
\end{equation}
Although $J^u$ and $J^u_1(\alpha_0,\alpha_1)$ are not necessarily everywhere smaller than unity,
there exists $C,\lambda_+,\ \lambda_->0$ such that, 
for any $n>0$, for any $\bal$ with $|\bal|=n$,
\begin{equation}\label{e:decay}
C^{-1}\,\e^{-n(d-1)\,\lambda_{+}}\leq J^u_n(\bal)\leq C\,\e^{-n(d-1)\,\lambda_{-}}\,.
\end{equation}
One can take $\lambda_+=\lambda_{\max}(1+\dia)$, where $\lambda_{\max}$ is the maximal expanding 
rate in Theorem.~\ref{t:thetheorem}.
We now give our central estimate, easy to draw from \cite[Corollary~3.4]{AN07}.
\begin{thm}
\label{t:main} 
Fix small positive constants $\dia$, $\delta$, $\delta'$ and a constant $0<C_\delta<(2\delta)^{-1}-1$. 
Take an open cover $M=\bigcup_k\Omega_k$ of diameter $\leq \dia$ and
an associated quantum partition $\cP=(P_k)_{k=1}^K$. There exists 
$\hbar_{0}$ such that, for any $\hbar\leq \hbar_{0}$, 
for any positive integer $n\leq C_{\delta}|\log\hbar|$, and
any pair of sequences $\bal$, $\bal'$ of length $n$,
\begin{equation}\label{e:main}
\norm{P_{\bal\bal'}\Op(\chi^{(n)})}=
\norm{P_{\bal'}\, U_\hbar^{n}\, P_{\bal} \Op(\chi^{(n)})}
\leq C\,\hbar^{-\frac{d-1}2-\delta}\,\e^{n\delta}\,\sqrt{J^u_n(\bal)\,J^u_n(\bal')}\,.
\end{equation}
The constant $C$ only depends on the Riemannian manifold $(M,g)$. 
If we take $n=n_E$, this takes the form
\begin{equation}\label{e:main-n_E}
\norm{P_{\bal'}\, U_\hbar^{n_E}\, P_{\bal} \Op(\chi^{(n_E)})}
\leq C\,\hbar^{-\frac{d-1+c\delta}2}\, \sqrt{J^u_{n_E}(\bal)\,J^u_{n_E}(\bal')}\,,
\end{equation}
where $c=2+2\lambda_{\max}^{-1}$. 
\end{thm}
The idea of proof in Theorem \ref{t:main} is rather simple, although the technical 
implementation is cumbersome. 
We first show that for any normalized state $\psi$, the state  $\Op(\chi^{(n)})\psi$ can be 
essentially decomposed into a superposition of
$\hbar^{-d}|\supp\chi^{(n)}|$
normalized Lagrangian states, supported on Lagrangian manifolds transverse to the stable foliation. 
In fact the Lagrangian states we work with are
truncated $\delta$--functions,
supported on lagrangians of the form $\cup_t g^t S_z^* M$. The action of the operator
$U^{n}P_{\bal\bal'}=P_{\alpha'_{n-1}}U\cdots U P_{\alpha_0}$ on such Lagrangian
states can be analyzed through WKB methods, and is simple to understand at the classical level~: 
each application of the propagator $U$ stretches the Lagrangian along
the unstable direction (the rate of stretching being described by the local unstable Jacobian), 
whereas each operator $P_k$ ``projects'' on a piece of Lagrangian of diameter $\dia$. 
This iteration of stretching and cutting accounts for the exponential decay. 
The $\bal\bal'$-independent factor on the right of \eqref{e:main-n_E}
results from adding together the contributions of all the initial Lagrangian states. 
Notice that this prefactor is
smaller than in Theorem.~\ref{t:mainestimate} due to the condition $C_\delta<(2\delta)^{-1}-1$.

\begin{rem}\label{r:improv}
In \cite{AN07} we used the same quantum partition $\cP^{\vee n_E}$ for $\pi$ and $\tau$ in 
Theorem.~\ref{t:WEUP}. As a result, we needed to estimate from above the norms
$\norm{P_{\bal'}^*\,U^{n_E}\,P_{\bal}\Op(\chi^{(n_E)})}$ (see  \cite[Theorem.~2.6]{AN07}). The proof
of this estimate was much more involved than the one for \eqref{e:main-n_E}, since it required to
control long pieces of unstable manifolds. By using instead the two partitions $\cP^{(n)}$, $\cT^{(n)}$,
we not only prove a more precise lower bound \eqref{e:main1} on the KS entropy, but also 
short-circuit some fine dynamical analysis.
\end{rem}

\subsubsection{Applying Theorem~\ref{t:WEUP}: Step 2}\label{s:weights}
There remains to choose
the weights $(v_{\bal},w_{\bal})$ to use in Theorem~\ref{t:WEUP}.
Our choice is guided by the following idea: 
in \eqref{e:c_C}, the weights should balance the variations (with respect to $\bal,\bal'$)
in the norms, such as to make all terms in \eqref{e:c_C} of the same order. 
Using the upper bounds \eqref{e:main-n_E}, we end up with the following choice for all
$\bal$ of length $n_E$:
$$
v_{\bal}= w_{\bal}\defeq J^u_{n_E}(\bal)^{-1/2 }\,.
$$
 
From \eqref{e:decay}, there exists $K_1>0$ such that, for $\hbar$ small enough, all the weights are 
contained in the interval $[1,\hbar^{-K_1}]$, as announced in \S\ref{s:mainapp}.
Using these weights, the estimate \eqref{e:main-n_E} implies the following bound 
on the coefficient \eqref{e:c_C}:
$$
\forall \hbar<\hbar_0,\qquad c_{\cO}^{\bv, \bw}(\cU)\leq C\,\hbar^{-\frac{d-1+c\delta}2}\,.
$$
We can now apply Corollary~\ref{c:WEUP} to the particular case of 
the eigenstates $\psi_\hbar$. We choose $L$ such that
$L-K_0-2K_1>-\frac{d-1+c\delta}2$, so from Corollary~\ref{c:WEUP} we draw the following
\begin{prop}\label{p:WEUP} 
Let $(\psi_\hbar)_{\hbar\to 0}$ be our sequence of eigenstates \eqref{e:eigenstate}.
In the semiclassical limit, the pressures of $\psi_\hbar$ satisfy
\begin{equation}\label{e:ineg}
p_{\cP^{\vee n_E}, \bv}(\psi_\hbar) + p_{\cT^{\vee n_E}, \bw}(\psi_\hbar)
\geq -\frac{(d-1+c\delta)\lambda_{\max}}{(1-\delta')}\; n_E +\cO(1)\,. 
\end{equation}
\end{prop}
If $M$ has constant curvature we have $\log J^n_{\bal}\leq - n (d-1)\lambda_{\max} (1-\cO(\dia))$ 
for all $\bal$ of length $n$, 
and the above lower bound can be written 
$$
h_{\cP^{\vee n_E}}(\psi_\hbar) + h_{\cT^{\vee n_E}}(\psi_\hbar)\geq  
(d-1)\lambda_{\max}\big(1+\cO(\dia,\delta,\delta')\big)\,n_E\,.
$$
As opposed to \eqref{e:lower-hn}, the above inequality provides a nontrivial
lower bound for the quantum entropies at the time $n_E$, which is smaller than
the time $n_1$ of \eqref{e:Ehrenf0}, and will allow to connect those entropies to the KS entropy
of the semiclassical measure (see below).

\subsubsection{Subadditivity until the Ehrenfest time}\label{s:subadd}

Even at the relatively small time $n_E$, the connection between 
the quantum entropy $h(\psi_\hbar,\cP^{\vee n_E})$ and the classical 
$h(\mu,\cP_{sm}^{\vee n_E})$ is not completely
obvious: both are sums of a large number of terms ($\asymp \hbar^{-K_0}$).
Before taking the limit $\hbar\to 0$, we will prove
that a lower bound of the form \eqref{e:ineg} still holds if we replace $n_E \asymp |\log\hbar|$ 
by some fixed $n_o\in\IN$,
and $\cP^{\vee n_E}$ by the corresponding quantum partition $\cP^{\vee n_o}$. The link between
quantum pressures at times $n_E$ and $n_o$ is provided by
the following {\em subadditivity property}, which is the semiclassical analogue 
of the classical subadditivity of pressures for invariant measures (see \eqref{e:cl-subadd}).
\begin{prop} [Subadditivity]\label{p:subadd}
Let $\del'>0$. There is a function $R(n_o,\hbar)$, and a real number $R>0$ independent of $\delta'$,
such that, for any integer $n_o\geq 1$, 
$$
\limsup_{\hbar\To 0}\abs{R(n_o, \hbar)}\leq R
$$
and with the following properties. For any small enough $\hbar>0$, any integers $n_o$,
$n\in\IN$ with $n_o+n\leq n_E(\hbar)$, for any $\psi_\hbar$ normalized 
eigenstate satisfying \eqref{e:eigenstate}, the following inequality holds:
$$
p_{\cP^{\vee (n_o+n)},\bv}(\psi_\hbar)\leq 
p_{\cP^{\vee n_o},\bv}(\psi_\hbar)+p_{\cP^{\vee n},\bv}(\psi_\hbar) 
+R(n_o, \hbar)\,.
$$
The same inequality is satisfied by the pressures $p_{\cT^{\vee n},\bw}(\psi_\hbar)$.
 \end{prop}
To prove this proposition, one uses a refined version of Egorov's theorem \cite{BouzRob}
to show that the non--commutative 
dynamical system formed by $(U_\hbar^t)$ acting (through Heisenberg) on observables supported near 
$\cE$ is (approximately) commutative
on time intervals of length $n_E(\hbar)$. Precisely, we showed in \cite{AN07} that,
provided $\dia$ is small enough,
for any $a,b\in C^\infty_c(\cE^{\dia})$,
$$
\forall t\in [-n_E(\hbar),n_E(\hbar)],\qquad 
\norm{[\Op_\hbar(a)(t), \Op_\hbar(b)]}=\cO(\hbar^{c\delta'}),\qquad\hbar\to 0\,,
$$
and the implied constant is uniform with respect to $t$.
Within that time interval, the operators $P_{\alpha_j}(j)$ appearing in the 
definition of the pressures commute up to small semiclassical errors.
This almost commutativity explains why the quantum pressures $p_{\cP^{\vee n},\bv}(\psi_\hbar)$ satisfy 
the same subadditivity property as the classical entropy \eqref{e:cl-subadd}, for times
smaller than $n_E$.

Thanks to this subadditivity, we may finish the proof of Theorem.~\ref{t:thetheorem}.
Fixing $n_o$, using for each $\hbar$ the Euclidean division $n_E(\hbar)=q(\hbar)\,n_o+r(\hbar) $ 
(with $r(\hbar)< n_o$), Proposition~\ref{p:subadd} implies that
for $\hbar$ small enough,
$$
\frac{p_{\cP^{\vee n_E},\bv}(\psi_\hbar)}{n_E}\leq \frac{p_{\cP^{\vee n_o},\bv}(\psi_\hbar)}{n_o}
+\frac{p_{\cP^{\vee r},\bv}(\psi_\hbar)}{n_E}+\frac{R(n_o, \hbar)}{n_o}\,.
$$
The same inequality is satisfied by the pressures $p_{\cT^{\vee n},\bw}(\psi_\hbar)$.
Using \eqref{e:ineg} and the fact that $p_{\cP^{\vee r},\bv}(\psi_\hbar)$ 
stays uniformly bounded when $\hbar\to 0$, we find
\begin{equation}\label{e:bientotfini}
\frac{p_{\cP^{\vee n_o},\bv}(\psi_\hbar)+p_{\cT^{\vee n_o},\bw}(\psi_\hbar)}{n_o} \geq 
-\frac{2(d-1+c\delta)\lambda_{\max}}{2(1-\delta')} 
-\frac{2R(n_o, \hbar)}{n_o}+\cO_{n_o}(1/n_E)\,.
\end{equation}
We are now dealing with quantum partitions $\cP^{\vee n_o}$, $\cT^{\vee n_o}$, 
for $n_0\in\IN$ independent of $\hbar$. At this level the quantum and classical entropies
are related through the (finite time) Egorov theorem, as we had noticed in \S\ref{s:naive}.
For any $\bal$ of length $n_o$, the weights $\norm{P_{\bal}\,\psi_\hbar}^2$ and
$\norm{P_{\bal}^*\,\psi_\hbar}^2$ both converge to $\mu\big((\bbbone_{M_{\bal}}^{sm})^2\big)$,
where we recall that
$$
\bbbone_{M_{\bal}}^{sm}=(\bbbone_{M_{\alpha_{n_o-1}}}^{sm}\circ g^{n_o-1})\times \ldots\times
(\bbbone_{M_{\alpha_1}}^{sm}\circ g)\times \bbbone_{M_{\alpha_0}}^{sm}\,.
$$
Thus, both entropies $h_{\cP^{\vee n_o}}(\psi_\hbar)$, $h_{\cT^{\vee n_o}}(\psi_\hbar)$ 
semiclassically converge to the classical entropy
$h_{n_o}(\mu,\cP_{sm})$.
As a result, 
the left hand side of \eqref{e:bientotfini} converges to
\begin{equation}\label{e:pressure0}
2\,\frac{ h_{n_o}(\mu,\cP_{sm})}{n_o}+\frac2{n_o}
\sum_{|\bal|=n_o}\mu\big((\bbbone_{M_{\bal}}^{sm})^2\big)\;\log J^u_{n_o}(\bal)\,.
\end{equation}
Since $\mu$ is $g^t$-invariant and $J^u_{n_o}$ has the multiplicative structure
\eqref{e:multi}, the second term in \eqref{e:pressure0} can be simplified:
$$
\sum_{|\bal|=n_o} \mu\big((\bbbone_{M_{\bal}}^{sm})^2\big)\,\log J^u_{n_o}(\bal)=
(n_o-1)\sum_{\alpha_0, \alpha_1}\mu\big((\bbbone_{M_{(\alpha_0,\alpha_1)}}^{sm})^2\big)\,
\log J^u_1(\alpha_0,\alpha_1)\,.
$$
We have thus obtained the lower bound
\begin{equation}\label{e:classicentropy}
\frac{ h_{n_o}(\mu,\cP_{sm})}{n_o} 
\geq -\frac{n_o-1}{n_o}\sum_{\alpha_0, \alpha_1}
\mu\big((\bbbone_{M_{(\alpha_0,\alpha_1)}}^{sm})^2\big)\,\log J^u_1(\alpha_0,\alpha_1)
-\frac{(d-1+c\delta)\lambda_{\max}}{2(1-\delta')}-\frac{R}{n_o}\,.
\end{equation}
At this stage we may forget about $\delta$ and $ \delta'$.
The above lower bound does not depend on the derivatives
of the functions $\bbbone_{M_{\bal}}^{sm}$, so the same bound carries over if we replace
$\bbbone_{M_{\bal}}^{sm}$ by the
characteristic functions $\bbbone_{M_{\bal}}$.
We can finally let $n_o$ tend to $+\infty$, then let the diameter $\dia$ tend to $0$. 
The left hand side converges to $h_{KS}(\mu)$ while,
from the definition \eqref{e:coarse-Jac},
the sum in the right hand side of \eqref{e:classicentropy} converges to the integral
$\int_{\cE} \log J^u(\rho)d\mu(\rho)$ as $\dia\to 0$, which proves \eqref{e:main1}.

$\hfill\square$

\end{document}